\documentstyle[12pt, epsf]{article}

\newcommand{\be}{\begin{equation}}
\newcommand{\ee}{\end{equation}}
\newcommand{\beqs}{\begin{eqnarray}}
\newcommand{\eeqs}{\end{eqnarray}}

\def\({\left(}
\def\){\right)}

\def\f{\frac}

\def\N{${\cal N}$}

\begin{document}
\begin{titlepage}

\begin{flushright}
\begin{tabular}{l} ITP-SB-99-36 \\ hep-th/9907142 \\ July, 1999
\end{tabular}
\end{flushright}

\vspace{8mm}
\begin{center} {\Large  Fundamental vs. Solitonic Description of D3 branes}

\vspace{20mm}

I.Y.~Park  \footnote{email: ipark@insti.physics.sunysb.edu},

\vspace{10mm} 

\vspace{2mm} 
C.N. Yang Institute for Theoretical Physics \\ 
State University of New York	\\ 
Stony Brook, N. Y. 11794-3840 \\

\vspace{20mm}

\begin{abstract}
Type IIB string theory expanded around D3 brane backgrounds describes the dynamics of D3 branes as solitonic objects. On the other hand, there is a fundamental description of them via Polchinski's open strings with Dirichlet boundary conditions. Since these two descriptions describe the dynamics of the same objects, D3 branes, it is natural to believe that they are dual. Therefore at this level, we have a string-string duality as opposed to a string-field theory duality. Once we take the same limits in both descriptions, Maldacena Conjecture in its weaker form follows. We try to make this viewpoint precise and study the implication of it for the stronger form of Maldacena Conjecture. 
\end{abstract}

\end{center}

\vspace{35mm}

\end{titlepage}
\newpage
\setcounter{page}{1}
\pagestyle{plain}
\pagenumbering{arabic}
\renewcommand{\thefootnote}{\arabic{footnote}} \setcounter{footnote}{0}

   In the progress of string theory during the past several years, D-branes \cite{P} have played a central role. In particular, the authors of \cite{HK,K1,GKT,GK} studied the absorption of scalars by D3 branes and noticed that supergravity calculations of the cross sections for various scattering processes are reproduced by the gauge theories on the branes. More recently, by studying the extremal 3 brane configurations of IIB supergravity Maldacena made a conjecture \cite{Mal}: The weaker form of the conjecture states that IIB supergravity on AdS$_5\times $S$_5$ is holographically \cite{T,S} dual to large N, ${\cal N}$=4, d=4 Super Yang-Mills (SYM$_4$) theory and the stronger form states that IIB {\em string} theory on AdS$_5\times$S$_5$ is dual to SYM$_4$ for any value of $g_s$ and N. There are several reviews of the topic, for example, \cite{IK,DRP,Pet,AGMOO}.\\
\indent Large amount of literature has appeared so far to support the weaker form of the conjecture. However, not much has been done along the line of proving or deriving the conjecture. In this article, we address the question of why the weaker form of the conjecture should be true. Similar reasoning appeared in \cite{Kreasoning}. However the viewpoint we take does not lead to the stronger form. Instead we seem to obtain a string-string duality. More specifically, we propose and try to make precise the following: IIB string theory on AdS$_5\times$S$_5$ should be dual to oriented open string theory\footnote{The closed strings produced by open string interactions are understood throughout the article.} propagating in the same background.\\    

\indent A couple of years ago, it was shown \cite{DKPS} that closed strings (or supergravity in the low energy) describe the dynamics of D-branes at length scales much larger than the string scale while it is the open strings (or world volume theory in the low energy) which dominate at the sub-stringy scales. Then one may ask whether one can put together open strings, closed strings and branes in one theory such that the theory is valid for the entire range of length scale.\\
\indent This is a very natural question because D-branes are, in some sense, topological defects that convert closed strings into open strings when the close strings touch the branes \cite{HK,K1,GKT,GK,K2}. Then there should be reversed processes in which open strings become closed strings and leave the branes. We take these properties of D-branes as our axiomatic starting point: The ``physical'' system under consideration consists of D-branes, open strings and closed strings, and we look for theories that describe their dynamics. Below we argue that there are (at least) two ways to combine these three objects (although there are some subtleties) and they correspond to two different stringy descriptions of D3 brane\footnote{In this article, we only consider D3 branes which are non-singular \cite{GT}. For Dp branes with p$\neq$3, there may be subtleties with the singularities.}  dynamics: one is ``fundamental'' and the other is ``solitonic''. In the solitonic approach, one looks for a theory in which the D-branes are embedded as solitons while one would consider the coordinates of the branes as dynamical fields in the fundamental description. For convenience and reasons which will become more clear, we will use the same word ``embedding'' for both descriptions.\\              
\indent It is well known that type IIB supergravity has extremal solutions of parallel 3-brane configurations \cite{HS,DKL}. For simplicity we only consider N coincident D-branes in this article. The solutions preserve one half of the supersymmetry and carry Ramond-Ramond (RR) charges. These facts allow one to view these particular solutions as a low energy realization of the D3 branes introduced by Polchinski. Even with the full string corrections, one expects to have corresponding solutions: a D3 brane will be a soliton of the full IIB string theory. Type IIB string theory expanded around them provides a solitonic description of D3 brane dynamics, Fig 1(a). In this description, D3 branes appear as solutions of field equations. They interact with each other by exchanging the perturbative spectrum of IIB theory excited ``around'' them. The situation is very similar to that of monopoles in ordinary field theories, where one considers perturbative excitations around a monopole solution.\\
\indent Then the subtlety lies in the way the open strings appear in this description. In other words, if IIB string theory expanded around a D3 brane solution correctly describe the dynamics (we assume it does), how does it incorporate the open strings? To answer this question precisely will require much work. However, given the fact that the open strings are associated with the excitations of the branes, a natural guess will be that the combined system of branes and open strings on them should be somehow associated with solutions with various non-extremality. We will return to this matter later. In summary, we started with  IIB closed stings, the D branes appear as solitons, open string effects will be embedded in some solutions with energies higher than those of extremal solutions and IIB string theory expanded around the D3 brane solutions describes their dynamics as solitons.\\ 
\begin{figure}[!ht]
\centerline{
        \begin{minipage}[b]{12cm}
                \epsfxsize=12cm
                \epsfbox{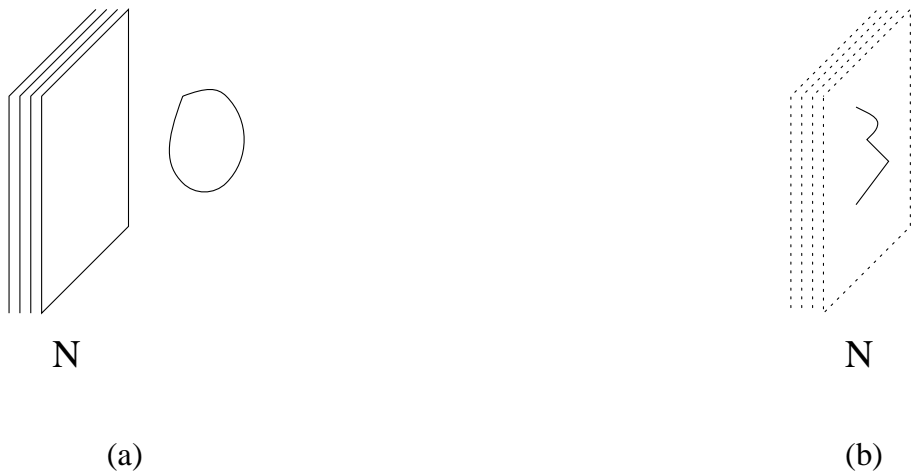}
        \end{minipage}
}   
\caption{(a) D3 branes appear as backgrounds in the solitonic description. (b) D3 branes appear as Dirichlet boundary conditions (represented by the dotted lines) in the fundamental description. }     
\label{fig:2all5}
\end{figure}         
 \indent  Now we seek another way of putting together the open strings, closed strings and D3 branes. Since we started with closed strings above, we start with D3 branes this time. One might consider the supermembrane Lagrangian \cite{BST} introducing the coordinates for the branes and try to quantize the theory. However, a direct quantization of such extended objects is known to be a difficult task \cite{WLN,deW}. Quantization of bosonic membranes has been discussed more recently in \cite{Kaku}. Fortunately one can achieve the same goal by an alternative description through open strings to which we now turn.\\
\indent Obviously the open strings must be oriented and have Dirichlet boundary conditions since the open strings we need are exactly the same open strings produced when the closed strings touch the D-branes. In this description, D branes are introduced as boundary conditions, Figure 1(b). (D branes are not merely boundary conditions, but they should provide a background in which open strings propagate. We will discuss this point later.) Finally closed strings are produced through the interactions of open strings, which will describe the reversed processes mentioned previously.\\
\indent It has been argued by Witten \cite{W2} that in the low energy, the dynamics of D3 branes are described by SYM$_4$ theory. SYM$_4$ has scalar fields which have the interpretation of the transverse coordinates\footnote{There may be higher derivative corrections to the identification of the scalars and the transverse coordinates of the branes \cite{GKPR}.}  of the D-branes. Therefore the open string description can be considered as a fundamental description of D branes. It is a {\em fundamental} description since the coordinates of the branes appear in the Lagrangian as dynamical variables.\\
\indent Above we have found two different ways of combining the three objects: open strings, closed strings and D3 branes. These in turn correspond to two different descriptions of the same D3 brane system: one is a solitonic description of IIB string theory for the D3 brane configuration under consideration, and the other is an open string description with Dirichlet boundary conditions. This is exactly the concept of duality: the two descriptions must be dual since they describe the same D3 brane system, which must have unique physics, at the same energy scale, the string scale. Therefore before taking any limits, we have a string-string duality. We will call this ``fundamental-solitonic duality''. In the solitonic description, closed strings are ``primary'' objects while the open strings are associated with some excited states of 3-brane configurations. On the other hand, it is the open strings which are primary in the fundamental description and the closed strings appear through the interactions of the open strings. Below we take this viewpoint seriously and  study the implications of it.\\ 
\indent The organization of this paper is as follows: First, we briefly demonstrate that the weaker form of Maldacena conjecture follows from the fundamental-solitonic duality by taking the same limits on both sides of the two descriptions. Then we argue that in high energy, one should expect a string-string duality rather than the stronger form of the Maldacena conjecture which is a string-field theory duality. These two points can be made without finding the precise statement of the fundamental-solitonic duality. Then we try to make this string-string duality more precise and study the additional implications of it. Finally we conclude with a summary and open problems.

\begin{figure}[!ht]
\centerline{
        \begin{minipage}[b]{10cm}
                \epsfxsize=10cm
                \epsfbox{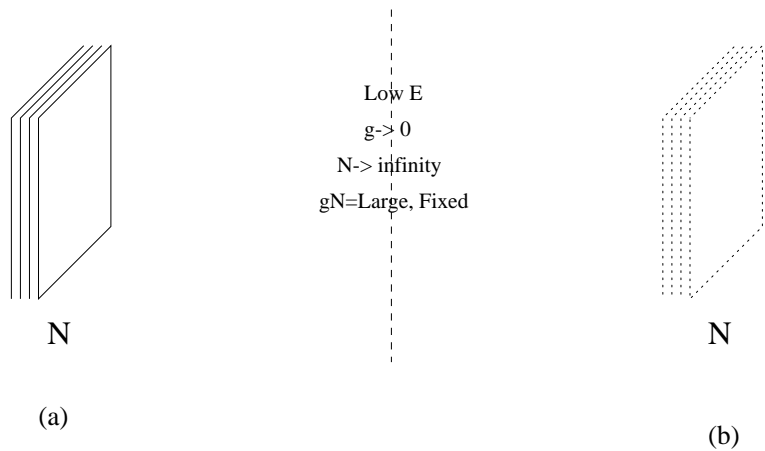}
        \end{minipage}
}
\caption{We take common limits in Solitonic Description, (a) and Fundamental Description, (b)}     
\label{}
\end{figure}                                  
             
To show that our viewpoint leads to the weaker form of Maldacena Conjecture, we consider N coincident D3 branes, Figure 2. Once again, it should be emphasized that we take limits {\em starting from the string-string duality}. Other than the fact that we do not take any limit for ``$r$'', the coordinate transverse to the branes\footnote{We will discuss the near-horizon limit of $r$ later.}, the argument should be rather familiar and we will be brief. Figure 2(b) represents the fundamental description and Figure 2(a) the solitonic one. Consider the low energy limit where one can neglect the massive string modes: the effective actions of both sides contain only {\em massless} string modes. We also impose the following common limits\footnote{Similar limits were taken in \cite{K1}.} on both sides, N $\rightarrow\infty ,  g_s \rightarrow 0, g_sN=$fixed but very large.  First consider the description (b), the fundamental description. According to Witten \cite{W2}, the low energy effective action of N coincident D3 branes is U(N) gauge theory. Even though $g_s\rightarrow 0$, we can not neglect the quantum corrections (because the effective loop expansion parameter is $g_sN$), but rather the theory becomes strongly coupled. Therefore the resulting theory in the description (b) is ${\cal N}$=4, d=4 U(N) SYM with quantum correction.  \\    
\indent This is in contrast with the solitonic description where the theory becomes classical. The metric configuration for N coincident D3 branes is \cite{HS}
\beqs
ds^2 &=&H^{-\f{1}{2}}dx_4^2+H^{\f{1}{2}}\left(dr^2+r^2d\Omega_5^2\right) \nonumber \\
  H  &=& 1+\f{4\pi g_s N {{\alpha}^{\prime}}^2}{r^4} \label{metric}
\eeqs

\noindent In the limit where $g_sN$ is very large, the geometry of the solitonic description becomes AdS$_5\times$S$_5$.  The radius of curvature of S$_5$, $R$, is given by $R^2={\alpha}^\prime\sqrt{4\pi g_s N }$. The massive stringy corrections are also neglected since they are expected to come as a geometrical series of $\f{{\alpha}^{\prime}}{R^2}=\f{1}{\sqrt{4\pi g_sN}}$ which becomes negligibly small in the limit we are taking. Therefore we have classical supergravity on AdS$_5\times$S$_5$.\\
\indent Since we started with two dual descriptions, the limiting theories must be dual too: type IIB supergravity on AdS$_5\times$S$_5$ is holographically dual to large N strongly coupled SU(N)\footnote{It has been argued \cite{W1,AW} that the bulk theory corresponds to SU(N) part of the U(N) gauge group.} SYM$_4$ and the precise form of the connection between the two theories was proposed in \cite{GKP,W1}.\\
\indent Now we turn to the implication of our viewpoint for the stronger form of the Conjecture. It is necessary to keep only the massless modes of open string description to obtain the SYM. Since we must impose the same limit for the solitonic description, it seems to be impossible to have full IIB {\em string} theory with all the massive modes if our viewpoint is correct. In other words, without taking the limits discussed above one will have oriented open {\em string} theory as a theory dual to IIB string theory. Therefore, instead of the stronger Conjecture, we propose below to consider oriented open string theory\footnote{Extension of the weaker form of Maldacena Conjecture to a duality between supergravity and DBI action was discussed in \cite{GHKK,deA}.} as a theory dual to IIB string theory on AdS$_5\times$S$_5$. 

  Having argued that the fundamental-solitonic duality implies AdS$_5$/CFT$_4$ duality in its weaker form, we now turn to the more precise statement of this string-string duality itself and study other implications of it. For this, we consider Fig 3. From now on, we do not take any limits of $g_s, {\alpha}^{\prime}$ and N unless otherwise specified.

\begin{figure}[!ht]
\centerline{
        \begin{minipage}[b]{12cm}
                \epsfxsize=12cm
                \epsfbox{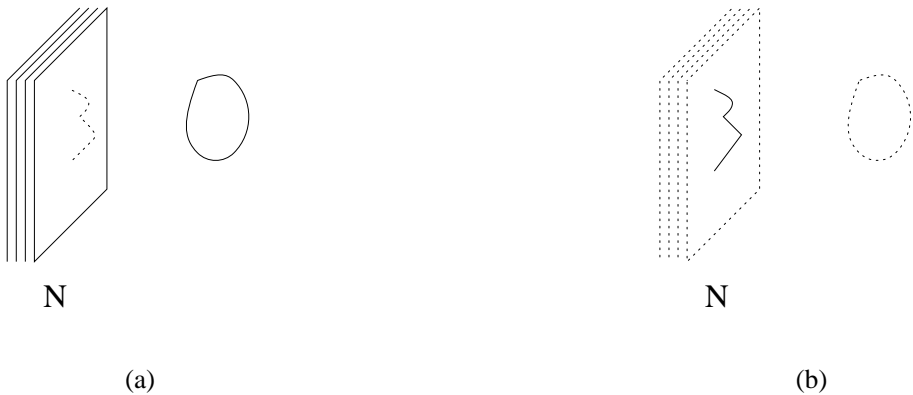}
        \end{minipage}
}
\caption{(a) Solitonic Description. (b) Fundamental Description}     
\label{}
\end{figure}                     

 The solitonic description, i.e., type IIB string theory on a background of N coincident D3 branes with RR charge \cite{BL}, is complicated at the technical level, given the fact that even the string theory on a simpler background such as AdS$_5\times$S$_5$ is not completely understood although there has been some progress \cite{Pe,KR,KT,Pol}. The D3 branes appear as backgrounds interacting each other by exchanging IIB strings, Figure 3 (a). It is not very clear how this description incorporates the open strings precisely. In \cite{DM,GKPe}, calculations of entropy of D-brane system using statistical mechanics of massless open string gas agreed with the Bekenstein-Hawking entropy of near-extremal solutions of supergravity. Therefore a natural guess would be that the massive open strings should be associated with non-extremal solutions with large non-extermality. In other words, when closed strings are incident on D-branes, open strings appear on the branes, but in the IIB description the combined system of the branes and the open strings on them may appear as some excited states and whether the open strings are massless or massive will be associated with the size of the non-extremality parameter, typically called $r_o$. Reversely, a brane in its excited state will decay emitting a closed string. \\
\indent To go to the fundamental description, we turn to our axiomatic starting point: When a closed string touches a D brane, it becomes an open string. This is the open string we must consider to obtain the fundamental description. In this description, the branes are embedded as Dirichlet boundary conditions, Fig 3(b). Since the branes are oriented, the open strings must be  oriented\footnote{There has been some recent discussion of oriented open and closed string theory in the frame work of string field theory \cite{Z}.}. Open string interactions will produce closed strings which must be IIB closed strings. The end points of open strings are constrained to move in a flat hyper-plane\footnote{In the low energy limit where we have a very poor resolution, the open strings on the flat hyper-plane will appear as {\em particles} moving in four dimensional Minkowski space (the flat hyper-plane) and this is the space where SYM theory is defined.}. However, the ``body'' of them propagates in the background produced by D branes. Therefore the D branes are not merely the boundary conditions but they also embed themselves as a background geometry for the open strings to propagate in.

We have just argued that IIB theory on D3 brane backgrounds must be dual to open string theory on the same background. Now we return to the near horizon limit $r \ll R $. This limit allows one to drop 1 in $H$ of eq (\ref{metric}) (without taking any limits for $g_s$ and $N$), and the metric becomes that of AdS$_5\times$S$_5$ again. Therefore we have IIB string theory on AdS$_5\times$S$_5$ for the solitonic description. Then we have to consider the same background for the open string description: we have open strings propagating in AdS$_5\times$ $S_5$. Therefore it is natural to believe that IIB string theory on AdS$_5\times$S$_5$ with N units of flux of 5-five form field strength should be dual to an open string theory in the same background\footnote{ It has been argued \cite{DT,FLZ} that the Dirac-Born-Infeld plus Wess-Zumino action should be expanded around AdS$_5$ rather than a flat background to obtain the dimension six operator predicted by supergravity \cite{KRV,R,DT}. This is natural according to our viewpoint.}. Then one immediate question is whether the open string description has the same symmetries as the IIB description. At the level of DBI action plus WZ term, superconformal symmetry has been discussed in \cite{CKV}. We expect to have superconformal symmetry at the level of full string theory too. After all, the symmetry should be viewed as coming from the background in the open string description too as it is true in the solitonic description.\\

\section*{Conclusion}
\indent In this article, the ``physical'' system of N coincident D3 branes, open strings and closed strings was considered. We argued that there are (at least) two ways to embed the system in string theory. The two ways correspond to two different descriptions of D3 branes: fundamental and solitonic. Since they are the descriptions of the same system, it is natural to believe that the two descriptions are dual. Taking appropriate, common limits on both sides of the descriptions leads to the weaker form of Maldacena Conjecture. However, we argued that in high energy, it is more natural to expect a string-string duality than the stronger form of Maldacena conjecture which takes the form of a string-field theory duality. In particular, we propose that IIB string theory on AdS$_5\times $S$_5$ with N units of the flux of five-form field strength should be dual to oriented open string theory in the same background. \\
\indent There are many details to fill in. Many of them are related to the poor status of string theory in various curved backgrounds. In particular, it will be very helpful to learn about oriented open string theory in the geometry of D3 branes or AdS$_5\times$S$_5$. It will also be interesting to study how precisely IIB description incorporates open string effects. We hope to make progress along these lines in near future.

\section*{Acknowledgments}
I thank J. Maldacena, H. Oogury and J. Schwarz for conversations and S.R. Das,  S. de Alwis, J. Polchinski and B. Zwiebach for communications. I am very grateful to I.R. Klebanov for reading the draft and making useful suggestions and corrections. I thank P. van Nieuwenhuizen for raising interesting questions. Finally I would like to thank I. Chepelev, B. Kulik, M. Ro\u{c}ek and specially F. Gonzalez-Rey for very helpful discussions and corrections.

\end{document}